\documentstyle[a4,11pt]{article}
\begin{document}
                                                                                
\setlength{\baselineskip}{16pt}
                                                                                
\begin{center}

{\bf \Large Controversy on a  dispersion relation for MHD waves}

\vspace{5mm}

Suresh Chandra\footnote{Visiting Associate of the IUCAA, Pune 411 007, India} 
and B.K. Kumthekar

\vspace{3mm}

\noindent
School of Physical Sciences, S.R.T.M. University,\\ Nanded 431 606, India\\
(email: suresh492000@yahoo.co.in)

\end{center}

\begin{abstract}
Kumar et al. (2006) obtained a fifth order polynomial in $\omega$ for the 
dispersion relation and pointed out that the calculations preformed  by Porter 
et al. (1994) and by Dwivedi \& Pandey (2003) seem to be in error, as they 
obtained a sixth order polynomial. The energy equation of Dwivedi \& Pandey
(2003) was dimensionally wrong. Dwivedi \& Pandey (2006) corrected the energy 
equation and still claimed that the dispersion relation must be a sixth order 
polynomial. The equations (11) $-$ (19) of  Dwivedi \& Pandey (2006) and the
equations (24) $-$ (32) Kumar et al. (2006) are the same. This fact has been 
expressed by Kumar et al. (2006) themselves. Even then they tried to show this 
set of equations on one side gives the sixth order polynomial as they got; on 
the other side, the same set of equations gives the fifth order polynomial as 
Kumar et al. (2006) obtained. The situation appears to be non-scientific, as 
the system of equations is a linear one. These are simple 
algebraic equations where the variables are to be eliminated. However, it is a 
matter of surprise that by solving these equations, two scientific groups are 
getting polynomials of different degrees. In the present discussion, we have 
attempted to short out this discrepancy. 
\end{abstract}

\section{Introduction}

For application of magnetohydrodynamics (MHD) in solar physics as well as in
plasma physics, dispersion relation, where $\omega$ is expressed as a function
of $k$, plays a key role. A controversy for the degree of the polynomial in 
$\omega$ for dispersion relation appeared  when Kumar et al. (2006, henceforth 
KKS) raised a point about the degree of the polynomial obtained by Porter et al.
(1994, henceforth PKS) and by Dwivedi \& Pandey (2003). Consequently, the 
results of PKS as well as of Dwivedi \& Pandey (2003) were kept before a 
question mark. Energy equations of Dwivedi \& Pandey (2003) was found erroneous 
(Klimchuk et al., 2004). After making correction in their energy equation, 
Dwivedi \& Pandey (2006, henceforth DP) made an attempt to show that the 
dispersion relation must be a sixth order polynomial. Since the results of an 
investigation involving MHD depend on the dispersion relation, it is important 
to resolve this controversy. This communication is an attempt to show that for 
the basic equations considered by DP, PKS and KKS, the dispersion relation 
comes out to be the same. In each case, it is a fifth order polynomial having 
the same coefficients.

\section{Basic equations of DP}
The basic equations used by DP as well as PKS are
\begin{eqnarray}
\frac{\partial \rho}{\partial t} +\nabla . \ (\rho \hspace{-0.5mm}
\stackrel{\rightarrow}{v} ) = 0 \hspace{4.5cm}\label{eq:int1}  \\ 
\rho \frac{D \hspace{-1mm} \stackrel{\rightarrow}{v}}{D t} =
 -\nabla p + \frac{1}{4 \pi}(\nabla \times \stackrel{\rightarrow}{B})\times
\stackrel{\rightarrow}{B} - \nabla . \Pi \hspace{1.2cm} \label{eq:int2} \\ 
\frac{D \hspace{-1mm} \stackrel{\rightarrow}{B}}{D t} = \nabla
\times (\stackrel{\rightarrow}{v} \times \stackrel{\rightarrow}{B}) 
\hspace{4.2cm} \label{eq:int3} \\ 
\frac{D p}{D t} + \gamma p (\nabla . \stackrel{\rightarrow}{v}) = (\gamma -1)
\big[Q_{th} + Q_{vis} - Q_{rad}\big]  \label{eq:int4} \\ 
 p = \frac{2\rho k_B T}{m_p} \hspace{5.7cm}  \label{eq:int5}
\end{eqnarray}

\noindent
with $Q_{th} = \nabla . \ \kappa \nabla T $ and $\frac{D }{D t} = \frac{\partial
 }{\partial t} + (\stackrel{\rightarrow}{v} . \nabla)$. These equations are, 
respectively,  the equation of continuity, equation of momentum, induction 
equation, energy equation and the equation of state. Here, $\rho$, $\stackrel{
\rightarrow}{v}$, $k_B$, $m_p$, $p$, $\stackrel{\rightarrow}{B}$, $\gamma$, $T$ 
and $\Pi$ are, respectively, the total mass density, velocity, Boltzmann 
constant, proton mass, total pressure, magnetic field, ratio of the specific 
heats, temperature and the viscous stress tensor. 

For small perturbations from the equilibrium (PKS, KKS):
\begin{eqnarray}
\rho = \rho_0 + \rho_1 \hspace{2.5cm} \stackrel{\rightarrow}{v} = \stackrel{
\rightarrow}{v}_1 \hspace{3.5cm} \stackrel{\rightarrow}{B} = \stackrel{
\rightarrow }{B}_0 + \stackrel{\rightarrow}{B}_1 \nonumber\\ 
p = p_0 + p_1 \hspace{2.5cm} T = T_0 + T_1 \hspace{2.5cm} \Pi = \Pi_0 + \Pi_1
  \nonumber
\end{eqnarray}

\noindent
where the equilibrium part is denoted by the subscript $``0"$ and the
perturbation part by the subscript $``1"$. For the magnetic field taken along
the $z$-axis, ({\it i.e.,} $\stackrel{\rightarrow}{B}_0 = B_0 \hat{z}$) and the
propagation vector $\stackrel{\rightarrow}{k} = k_x \hat{x} + k_z \hat{z}$, the
equations (\ref{eq:int1}) $-$  (\ref{eq:int5}) can be linearized in the
following form.
\begin{eqnarray}
\frac{\partial \rho_1}{\partial t} + \rho_0 (\nabla . \stackrel{\rightarrow}
{v}_1 ) = 0 \hspace{3.8cm} \label{eq:int6}\\ 
\nonumber  \\
\rho_0 \frac{\partial \hspace{-0.5mm} \stackrel{\rightarrow} v_1}{\partial t} =
-\nabla p_1 + \frac{1}{4 \pi} (\nabla \times \stackrel{\rightarrow}{B_1}) \times
 \stackrel{ \rightarrow}{B_0}- \nabla . \Pi_0  \label{eq:int7}\\ 
\nonumber  \\
\frac{\partial \hspace{-1mm} \stackrel{\rightarrow}{B_1}}{\partial t} = \nabla
\times (\stackrel{\rightarrow}{v_1} \times \stackrel{\rightarrow}{B_0})
\hspace{3.7cm} \label{eq:int8}\\ 
\nonumber  \\
\frac{\partial  p_1}{\partial t}+ \gamma p_0 (\nabla . \stackrel{\rightarrow}
{v}_1) + (\gamma -1) \kappa_\parallel k_z^2 T_1 = 0 \hspace{10mm}
\label{eq:int9}\\ \nonumber  \\
\frac{p_1}{p_0} = \frac{\rho_1}{\rho_0}+ \frac{T_1}{T_0}\hspace{5.0cm}
\label{eq:int10}
\end{eqnarray}
                                                                                
\noindent
For the perturbations that are proportional to $\mbox{exp} [i (\stackrel{
\rightarrow}{k} \hspace{-0.5mm} . \hspace{-0.5mm} \stackrel{\rightarrow}{r}
\hspace{-0.5mm} - \hspace{0.5mm} \omega t)]$, equations (\ref{eq:int6}) $-$  
(\ref{eq:int10}) reduce to the following algebraic equations
\begin{eqnarray}
\omega \rho_1 - \rho_0(k_x v_{1 x} + k_z v_{1 z}) = 0 \hspace{6.2cm} 
\label{eq:int11}\\
\omega\rho_0 v_{1 x}-k_x p_1-\frac{B_0}{4\pi}(k_x B_{1 z}-k_z B_{1 x})+
\frac{i\eta_0}{3}(k_x^2 v_{1 x}-2k_x k_z v_{1 z}) = 0 \label{eq:int12}\\
\omega\rho_0 v_{1 y} +\frac{B_0}{4\pi}(k_z B_{1 y}) = 0\hspace{6.9cm}
\label{eq:int13}\\
\omega\rho_0 v_{1 z}-k_z p_1+\frac{i\eta_0}{3}(4k_z^2 v_{1 z}-2k_x k_z v_{1 x})
= 0 \hspace{3.7cm}\label{eq:int14}\\ 
\omega B_{1 x} + k_z B_0 v_{1 x} = 0 \hspace{7.7cm}\label{eq:int15}\\
\omega B_{1 y} + k_z B_0 v_{1 y} = 0 \hspace{7.7cm}\label{eq:int16}\\
\omega B_{1 z} - k_x B_0 v_{1 x} = 0 \hspace{7.7cm}\label{eq:int17}\\
i\omega p_1 - i \rho_0 c_s^2 (k_x v_{1 x} + k_z v_{1 z}) - (\gamma-1)
\kappa_\parallel k_z^2 T_1 =0 \hspace{3cm} \label{eq:int18}\\ 
\frac{p_1}{p_0}-\frac{\rho_1}{\rho_0 }-\frac{T_1}{T_0} = 0 \hspace{8.1cm}
\label{eq:int19}
\end{eqnarray}

\noindent
where $c_s^2 = \gamma p_0/\rho_0$. Equations (\ref{eq:int13}) and 
(\ref{eq:int16}) for the variables $v_{1 y}$ and $B_{1 y}$ are decoupled from 
the rest and describe Alfv\'{e}n waves. The rest of the equations for $p_1$, 
$\rho_1$, $T_1$, $B _{1 x}$, $B_{1 z}$, $v_{1 x}$ and  $v_{1 z}$ describe 
damped magnetoacoustic waves. For  elimination of the variables $p_1$, $\rho_1$,
$T_1$, $B _{1 x}$, $B_{1 z}$, $v_{1 x}$ and  $v_{1 z}$, we have
\begin{eqnarray}
\begin{array}{cc}
\left|
\begin{array}{lllllll}
0 & \omega & 0 & 0 & 0 & -\rho_0 k_x & -\rho_0 k_z \\
-k_x &  0 & 0 & \frac{B_0}{4 \pi}k_z & -\frac{B_0}{4 \pi}k_x & \big(\omega
\rho_0+ \frac{i \eta_0}{3} k_x^2 \big) & -\frac{ 2 i \eta_0}{3} k_z k_x \\
-k_z &  0 & 0 & 0 & 0 &  -\frac{ 2 i \eta_0}{3} k_z k_x &\big(\omega\rho_0+
\frac{4 i \eta_0}{3} k_z^2 \big) \\
0 & 0 & 0 & \omega & 0 & k_z B_0 & 0 \\
0 & 0 & 0 & 0 & \omega & -k_x B_0 & 0 \\
i\omega & 0 & -(\gamma-1) \kappa_ \parallel k_z^2 &  0 & 0 & -i \rho_0 c_s^2 k_x
 & -i \rho_0 c_s^2 k_z \\
\frac{1}{p_0} & -\frac{1}{\rho_0} & -\frac{1}{T_0} & 0 & 0 & 0 & 0 \\
\end{array}
\right|        & \hspace{-5mm} \begin{array}{c}
= 0 \\  \\ \\  \\ \\ \\ \\
\end{array}\\
\end{array} \label{eq:int20}
\end{eqnarray}

\section{Basic equations of KKS}

The basic equations used by KKS are
\begin{eqnarray}
\frac{\partial \rho}{\partial t} +\nabla . \ (\rho \hspace{-0.5mm}
\stackrel{\rightarrow}{v} ) = 0 \hspace{4.4cm}\label{eq:intk1}  \\
\rho \frac{D \hspace{-1mm} \stackrel{\rightarrow}{v}}{D t} =
 -\nabla p + \frac{1}{4 \pi}(\nabla \times \stackrel{\rightarrow}{B})\times
\stackrel{\rightarrow}{B} - \nabla . \Pi \hspace{1.2cm} \label{eq:intk2} \\
\frac{\partial  \hspace{-1mm} \stackrel{\rightarrow}{B}}{\partial t} = \nabla
\times (\stackrel{\rightarrow}{v} \times \stackrel{\rightarrow}{B})
\hspace{4.2cm} \label{eq:intk3} \\ 
\frac{D p}{D t}-\frac{\gamma p D \rho}{\rho D t} = (\gamma -1)\big[Q_{th} +
Q_{vis} - Q_{rad}\big]  \label{eq:intk4}\hspace{.5cm}\\
 p = \frac{2\rho k_B T}{m_p} \hspace{5.6cm}  \label{eq:intk5}
\end{eqnarray}
                                                                                
\noindent
Comparison of the two sets [equations (\ref{eq:int1}) $-$ (\ref{eq:int5}) and
equations (\ref{eq:intk1}) $-$ (\ref{eq:intk5})] show that there is difference 
on the left side in the induction and energy equations. For this set of 
equations (\ref{eq:intk1}) $-$ (\ref{eq:intk5}), after going through the same 
procedure as discussed in the preceding section, we get the equations (KKS)
\begin{eqnarray}
\omega \rho_1 - \rho_0(k_x v_{1 x} + k_z v_{1 z}) = 0 \hspace{6.3cm}
\label{eq:intk11}\\
\omega\rho_0 v_{1 x}-k_x p_1-\frac{B_0}{4\pi}(k_x B_{1 z}-k_z B_{1 x})+
\frac{i\eta_0}{3}(k_x^2 v_{1 x}-2k_x k_z v_{1 z}) = 0 \label{eq:intk12}\\
\omega\rho_0 v_{1 y} +\frac{B_0}{4\pi}(k_z B_{1 y}) = 0\hspace{6.9cm}
\label{eq:intk13}\\
\omega\rho_0 v_{1 z}-k_z p_1+\frac{i\eta_0}{3}(4k_z^2 v_{1 z}-2k_x k_z v_{1 x})
= 0 \hspace{3.7cm}\label{eq:intk14}\\
\omega B_{1 x} + k_z B_0 v_{1 x} = 0 \hspace{7.7cm}\label{eq:intk15}\\
\omega B_{1 y} + k_z B_0 v_{1 y} = 0 \hspace{7.7cm}\label{eq:intk16}\\
\omega B_{1 z} - k_x B_0 v_{1 x} = 0 \hspace{7.7cm}\label{eq:intk17}\\
i\omega p_1 - i \rho_1\omega  c_s^2  - (\gamma-1)
\kappa_\parallel k_z^2 T_1 =0 \hspace{5.3cm} \label{eq:intk18}\\
\frac{p_1}{p_0}-\frac{\rho_1}{\rho_0 }-\frac{T_1}{T_0} = 0 \hspace{8.0cm}
\label{eq:intk19}
\end{eqnarray}

\noindent
Obviously, equation (\ref{eq:int18}) is different from  (\ref{eq:intk18}). 
DP started from the set of equations (\ref{eq:int1}) $-$ (\ref{eq:int5}), but 
the results reported by them are as given here in equations (\ref{eq:intk11}) 
$-$ (\ref{eq:intk19}). It appears that DP did not derive their equations
(11) $-$ (19), but adopted directly from KKS. It is noticeable that the present
equations (\ref{eq:int11}) $-$ (\ref{eq:int19}) are not available in the paper 
of PKS. Equations  (\ref{eq:intk13}) and (\ref{eq:intk16}) for the variables 
$v_{1 y}$ and $B_{1 y}$ are decoupled from the rest and describe Alfv\'{e}n 
waves. The rest of the equations for $p_1$, $\rho_1$, $T_1$, $B _{1 x}$, $B_{1 
z}$, $v_{1 x}$ and  $v_{1 z}$ describe damped magnetoacoustic waves. For 
elimination of the variables $p_1$, $\rho_1$, $T_1$, $B _{1 x}$, $B_{1 z}$, 
$v_{1 x}$ and  $v_{1 z}$, we have 
\begin{eqnarray}    
\begin{array}{cc}
\left|
\begin{array}{lllllll}
0 & \omega & 0 & 0 & 0 & -\rho_0 k_x & -\rho_0 k_z \\
-k_x &  0 & 0 & \frac{B_0}{4 \pi}k_z & -\frac{B_0}{4 \pi}k_x & \big(\omega
\rho_0 + \frac{i \eta_0}{3} k_x^2 \big) & -\frac{ 2 i \eta_0}{3} k_z k_x \\
-k_z &  0 & 0 & 0 & 0 &  -\frac{ 2 i \eta_0}{3} k_z k_x &\big(\omega\rho_0+ 
\frac{4 i \eta_0}{3} k_z^2 \big) \\
0 & 0 & 0 & \omega & 0 & k_z B_0 & 0 \\
0 & 0 & 0 & 0 & \omega & -k_x B_0 & 0 \\
i\omega & -i\omega c_s^2 & -(\gamma-1) \kappa_ \parallel k_z^2 &  0 & 0 & 0 & 
0 \\
\frac{1}{p_0} & -\frac{1}{\rho_0} & -\frac{1}{T_0} & 0 & 0 & 0 & 0 \\
\end{array}
\right|        &  \hspace{-5mm} \begin{array}{c}
= 0 \\  \\ \\  \\ \\ \\ \\
\end{array}\\
\end{array} \label{eq:intk20}
\end{eqnarray}   
                                                                                
\section{Discussion and conclusion}

In order to resolve the controversy, now, we are left with two determinants 
(\ref{eq:int20}) and (\ref{eq:intk20}), which correspond to the sets of 
equations used by DP (also PKS) and KKS, respectively. It is interesting to 
find out that both these determinants reduce to the following common 
determinant:

\begin{displaymath}
\begin{array}{cc}
\left|
\begin{array}{lll}
k_x \omega   & \omega^2\rho_0+i\frac{\omega\eta_0}{3}k_x^2-v_A^2\rho_0k^2 
 & -\frac{2 i \omega\eta_0}{3}k_x k_z \\
&&\\
k_z      & -\frac{2 i \eta_0}{3}k_x k_z   &\omega\rho_0+\frac{4 i \eta_0}{3}
k_z^2   \\
&&\\
c_0 \omega - i\omega^2     & c_0 p_0 k_x -i \rho_0 c_s^2 k_x
 \omega   &  c_0 p_0 k_z - i \rho_0 c_s^2 k_z\omega \\
\end{array}
\right|        & \begin{array}{c}
= 0 \\  \\ \\  \\
\end{array}\\
\end{array}
\end{displaymath}

\noindent
where $c_0 = (\gamma -1) \kappa_\parallel k_z^2 T_0/p_0$ and $v_A = B_0/\sqrt{4
 \pi \rho_0}$. This determinant can be solved to get the following dispersion relation.
\begin{eqnarray}
\omega^5 + i A  \omega^4 - B \omega^3 - i C \omega^2 + D \omega + i E = 0
\nonumber       
\end{eqnarray}
                                                                                
\noindent
where
\begin{eqnarray}
A = c_0 + \frac{\eta_0}{3 \rho_0} (k_x^2 + 4 k_z^2) \hspace{7.4cm}  \nonumber \\
B = \frac{c_0 \eta_0}{3 \rho_0} (k_x^2 + 4 k_z^2) + (c_s^2 + v_A^2) k^2
\hspace{5.7cm} \nonumber \\
C = \frac{3\eta_0}{\rho_0} c_s^2 k_x^2 k_z^2 + \frac{c_0 p_0 k^2}{\rho_0} + 
v_A^2 c_0 k^2 + \frac{4 \eta_0 v_A^2 k_z^2 k^2}{ 3 \rho_0}
\hspace{3.3cm} \nonumber \\ 
D = \frac{3 c_0 p_0 \eta_0 k_x^2 k_z^2}{\rho_0^2} + \frac{4 \eta_0 c_0 v_A^2
k_z^2 k^2}{3 \rho_0} + v_A^2 c_s^2 k_z^2 k^2 \hspace{3.8cm} \nonumber \\ 
E = \frac{v_A^2 c_0 p_0 k_z^2 k^2}{\rho_0} \hspace{8.5cm}  \nonumber
\nonumber
\end{eqnarray}

\noindent
Hence, the dispersion relation obtained from both sets of the basic equations 
of DP (also PKS) and KKS is a fifth order polynomial in $\omega$. The 
coefficients obtained here are the same as obtained by KKS. It may finally be 
resolved that the dispersion relations derived by KKS for the basic set of
equations (\ref{eq:intk1}) $-$ (\ref{eq:intk5}) is correct.

Though both the sets of basic equations produce a common dispersion relation,
some points regarding the discrepancy between the induction and energy equations
of the two sets can be noted as the following.

In the induction equation (\ref{eq:int3}), the term $D \hspace{-1mm} \stackrel{
\rightarrow}{B}/D t$ can be expressed as
\begin{eqnarray}
\frac{D \hspace{-1mm} \stackrel{\rightarrow}{B}}{D t} = \frac{\partial
 \hspace{-1mm} \stackrel{\rightarrow}{B}}{\partial t} + (\stackrel{\rightarrow}
{v} . \nabla) \stackrel{\rightarrow}{B} \nonumber
\end{eqnarray}                                                                                 
\noindent
Linearization of this equation gives
\begin{eqnarray}
\frac{D \hspace{-1mm} \stackrel{\rightarrow}{B}_1}{D t} = \frac{\partial  
\hspace{-1mm} \stackrel{\rightarrow}{B}_1}{\partial t} + (\stackrel{\rightarrow}
 {v}_1 . \nabla) \stackrel{\rightarrow}{B}_1 \nonumber
\end{eqnarray}                                                                                 
\noindent
The second term on right side can be dropped as it is a product of two 
perturbations. Thus, we have
\begin{eqnarray}
\frac{D \hspace{-1mm} \stackrel{\rightarrow}{B}_1}{D t} = \frac{\partial 
\hspace{-1mm} \stackrel{\rightarrow}{B}_1}{\partial t}  \nonumber
\end{eqnarray} 

\noindent
and the induction equation in the two sets give the same final equations.

In the energy equation (\ref{eq:int4}), the term $\gamma p (\nabla . 
\stackrel{\rightarrow}{v})$ reduces to $- i \rho_0 c_s^2 (k_x v_{1 x} + k_z 
v_{1 z})$ whereas in the equation (\ref{eq:intk4}), the term $-(\gamma p/\rho)
(D \rho/ D t)$ reduces to $- i \rho_1\omega  c_s^2$. However, after the 
calculations, no difference is found in the expression for dispersion relation.

\begin{center}
{\bf Acknowledgments}
\end{center}

This work was done during the visit of the authors to the IUCAA, Pune, India. 
We are thankful to the Department of Science \& Technology, New
Delhi and the Indian Space Research Organization (ISRO), Bangalore for
financial support in the form of research projects.

\begin{center}
{\bf References}
\end{center}
\begin{description}

\item{}  Dwivedi, B.N. \& Pandey, V.S. 2003, Solar Phys., 216, 
59

\item{}  Dwivedi, B.N. \& Pandey, V.S. 2006, arXiv:astro-ph/0611249

\item{}  Klimchuk, J.A., Porter, L.J. \& Sturrock, P.A. 2004, 
Solar Phys., 221, 47
                                                                                
\item{}  Kumar, N., Kumar, P. \& Singh, S. 2006, Astron. Astrophys.  453, 1067
                                                                                
\item{} Porter, L.J., Klimchuk, J.A. \& Sturrock, P.A. 1994, 
Astrophys. J., 435, 482.

\end{description}     

\end{document}